\documentclass[letterpaper,titlepage,11pt]{article}
\pdfoutput=1
\usepackage[colorlinks=true,urlcolor=blue,citecolor=magenta]{hyperref}
\usepackage{amssymb,amsmath,amsfonts}
\usepackage{color}
\usepackage{graphicx}
\usepackage{array}
\usepackage{epsfig}
\usepackage{epstopdf}
\usepackage{caption}
\usepackage{subcaption}
\usepackage{epsfig}
\usepackage{epstopdf}
\usepackage{eso-pic}
\usepackage{tensor}
\usepackage{float}
\hypersetup{urlcolor=red}

\def\clock{{\count0=\time
          \divide\count0 60
          \ifnum\count0<10 0\fi\the\count0
          \multiply\count0 -60 \advance\count0 \time
          :\ifnum\count0<10 0\fi \the\count0
        }}
\newcommand{\timestamp}{{\small\vbox{\hbox{\tt\jobname.tex}
\hbox{\the\day/\the\month/\the\year, \clock}}}}

\newcommand{\beq}{\begin{equation}}
\newcommand{\eeq}{\end{equation}}

\newcommand{\tr}{\mathop{{\rm Tr}}}

\numberwithin{equation}{section}
\begin{document}

\begin{titlepage}
\ \ \vskip 1.8cm
\centerline{\Huge \bf Uniqueness of Black Holes with Bubbles} 
\vskip 0.6cm
\centerline{\Huge \bf in Minimal Supergravity} 

\vskip 1.3cm \centerline{\bf Jay Armas} \vskip 0.3cm
\let\thefootnote\relax\footnote{\textcolor{red}{\url{http://www.jacomearmas.com}}}
\begin{figure}[!ht]
\vskip 0.1cm
\centerline{\includegraphics[scale=0.5]{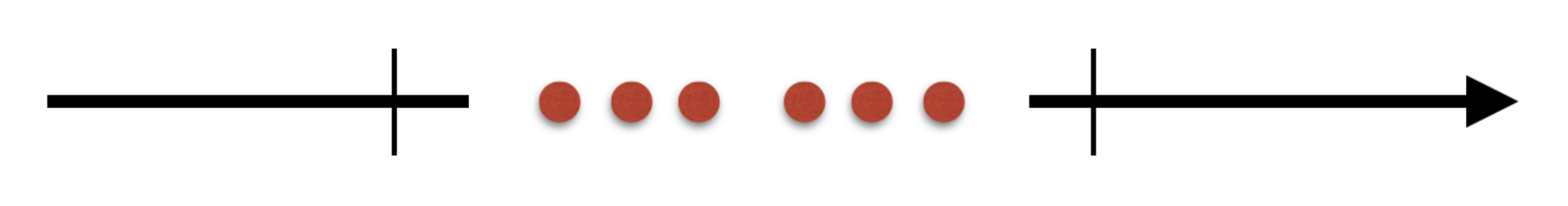} }	
\end{figure}

\begin{center}
\sl Albert Einstein Center for Fundamental Physics, University of Bern,\\
Sidlerstrasse 5, 3012 Bern, Switzerland
\end{center}
\vskip 0.3cm

\centerline{\small\tt jay@itp.unibe.ch}

\vskip 1.3cm \centerline{\bf Abstract} \vskip 0.2cm \noindent
We generalise uniqueness theorems for non-extremal black holes with three mutually independent Killing vector fields in five-dimensional minimal supergravity in order to account for the existence of non-trivial 2-cycles in the domain of outer communication. The black hole space-times we consider may contain multiple disconnected horizons and be asymptotically flat or asymptotically Kaluza-Klein. We show that in order to uniquely specify the black hole space-time, besides providing its domain structure and a set of asymptotic and local charges, it is necessary to measure the magnetic fluxes that support the 2-cycles as well as fluxes in the two semi-infinite rotation planes of the domain diagram.

%\vskip 0.5cm \leftline{\timestamp}
\end{titlepage}

%\pagestyle{empty}
%\small
%\begin{spacing}{1}
\tableofcontents
%\end{spacing}
%\tableofcontents
%\normalsize
%\newpage
\pagestyle{plain}
\setcounter{page}{1}

%%%%%%%%%%%%%%%%%%%%%%%%%%%%%%%%%%%%%%%%%%%%%%%%%%%%%%%%%%
\section{Introduction}
Recently, there has been an interest in black holes with non-trivial space-time topology outside the domain of outer communication as it was realised that in dimensions higher than four, space-like hyper surfaces may have non-trivial homology groups. This has triggered a series of papers where previous no-go theorems, uniqueness theorems and laws of black hole mechanics were put into question and in some cases successfully modified in order to account for the non-trivial space-time topology \cite{Gibbons:2013tqa, Kunduri:2013vka, Kunduri:2014iga}.

In five dimensions the non-trivial structure of the space-time is characterised by the existence of 2-cycles. In Einstein gravity coupled to a Maxwell field, a likely mechanism for supporting these 2-cycles (bubbles) is the magnetic flux. Gravitational solutions with sequences of such bubbles are of considerable interest as they may describe smooth soliton geometries \cite{Bena:2005va}, which, in the context of fuzzballs \cite{Mathur:2005zp, Mathur:2008nj}, are interpreted as black hole microstates. 

In a recent paper \cite{Kunduri:2014iga}, an example of an extremal black hole with non-trivial space-time topology and two rotational isometries was given in minimal supergravity and seems to provide the first counterexample to spherical black hole uniqueness in theories containing a Maxwell field. This note has the purpose of showing that, in minimal supergravity, if the black hole is non-extremal and has isometry group $\mathbb{R}\times U(1) \times U(1)$, then we can uniquely characterise it given its domain (or rod) structure \cite{Harmark:2004rm, Harmark:2005vn, Harmark:2009dh,Armas:2011ed}  and charges.

Uniqueness theorems in minimal supergravity have been written down for asymptotically flat spherical black holes \cite{Tomizawa:2009ua}, black rings \cite{Tomizawa:2009tb}, Lens-spaces and for black holes with multiple disconnected horizons \cite{Armas:2009dd}.\footnote{See \cite{Hollands:2012xy} for a review on uniqueness theorems.} In the latter case, it was shown that, in order to specify the black hole space-time uniquely it was necessary to define certain magnetic fluxes on spatial compact rods as well as to measure a set of local charges near each horizon. In the case of Lens-spaces, it was also realised in \cite{Armas:2009dd}, that the magnetic flux had to be measured in a semi-infinite rotation plane in the rod diagram.\footnote{In the context of Kaluza-Klein black holes in Einstein-Maxwell theory this was also realised in \cite{Yazadjiev:2010uu}.} In the context of asymptotically Kaluza-Klein black holes, a uniqueness theorem was also written down for black holes with a single connected horizon \cite{Tomizawa:2010xj}. However, all these uniqueness theorems \cite{Tomizawa:2009ua, Tomizawa:2009tb,Armas:2009dd, Tomizawa:2010xj} have not taken into account the possibility of non-trivial space-time topology\footnote{We note that in the case of an integrable sector of Einstein-Maxwell theory, non-trivial 2-cycles have been considered in \cite{Hollands:2007qf, Yazadjiev:2010uu}.}, and hence, of arbitrary large sequences of bubbles in the domain of outer communication.

In this note, we generalise the uniqueness theorems \cite{Tomizawa:2009ua, Tomizawa:2009tb,Armas:2009dd, Tomizawa:2010xj} in order to account for the possible 2-cycles present in the space-time. The necessary ingredients for such a generalisation were developed in \cite{Armas:2009dd} and here we apply them in a systematic way. For a single connected horizon component, we show that the magnetic fluxes on each spatial compact rod and spatial semi-infinite rod must be given in order for the solution to be uniquely specified. We also show that, in the case of the horizon having topology $S^{1}\times S^{2}$, the dipole charge which seemed to be necessary to show uniqueness of such solutions, can be replaced by the magnetic flux in a semi-infinite rotation axis. In the case of space-times with multiple disconnected horizons we analyse the most general class of solutions by considering an arbitrary arrangement of horizons and spatial compact rods. However, in this note, the uniqueness theorems proven in this case are not completely general and some constraints will have to be set on this arbitrary sequence of spatial and horizon rods. However, the uniqueness theorems cover all known black hole solutions.\footnote{We note that up to now, the uniqueness theorem proven in \cite{Armas:2009dd} applies to all known non-extremal black hole solutions with $\mathbb{R}\times U(1) \times U(1)$ isometry group in minimal supergravity.} 

This note is organised as follows. In Sec.~\ref{sec:review} we review the general method for proving uniqueness in the context of minimal supergravity based on a reduction to a non-linear sigma model. In Sec.~\ref{sec:domain} we define the domain (rod) structure of the most general solution. In Sec.~\ref{sec:single}, we show uniqueness of the most general solution with a single connected horizon. In Sec.~\ref{sec:multiple}, we generalise it to the case of multiple disconnected horizons and in Sec.~\ref{sec:kaluza} we extend it to asymptotically Kaluza-Klein space-times. Finally, in Sec.~\ref{sec:conclusions} we conclude.

%%%%%%%%%%%%%%%%%%%%%%%%%%%%%%%%%%%%%%%%%%%%%%%%%%%%%%%%%%
\section{Uniqueness theorems for Black Holes with Bubbles} \label{sec:review}
In this section we give a proof of the uniqueness of black holes with an arbitrary large sequence of non-trivial 2-cycles and multiple disconnected horizons in five-dimensional minimal supergravity. The proof follows closely the work of \cite{Tomizawa:2009ua, Tomizawa:2009tb,Armas:2009dd, Tomizawa:2010xj} which exploits the reduction of the theory to a non-linear sigma model. Because this proof extends these previous works, we refer to these for the details of this reduction and instead summarise the procedure. 

We assume that the black hole solutions are characterised by a set of three mutual independent Killing vector fields $V_{(0)}=\partial_t$, $V_{(1)}=\partial_\phi$ and $V_{(2)}=\partial_\psi$, where the first is associated with time translations and the other two with rotations, such that the isometry group of the solutions is $\mathbb{R}\times U(1) \times U(1)$. Assuming this set of symmetries allows us to reduce the theory characterised by the action 
\beq \label{eq:action}
S=\frac{1}{16\pi}\left(\int dx^5 \sqrt{-g}(R-\frac{1}{4}F^2) -\frac{1}{3\sqrt{3}}\int F\wedge F \wedge A\right)~~,
\eeq
where $F=dA$ is the field strength for the 1-form gauge field $A$, to a non-linear sigma model on a two-dimensional base space $\Sigma$ parametrised by the coordinates $(r,z)$ such that $\Sigma=\{(r,z)| r\ge0~,~-\infty<z<\infty\}$ \cite{Tomizawa:2009ua}. Due to the symmetry of the sigma-model, solutions of the theory \eqref{eq:action} can be collectively described by a set of sigma-model fields $\Phi^{A}$ defined on $\Sigma$, which form a symmetric unimodular matrix $\Theta$.  For the case of the theory \eqref{eq:action}, we have that $\Phi^{A}$ consists of the fields $\Phi^{A}=\{\lambda_{ab},\omega_a,\psi_a,\mu\}$ with $a=(\phi,\psi)$ which parametrise the metric and the gauge field according to
\beq \label{eq:metric}
ds^2=\lambda_{ab}(dx^a+a^{a}_tdt)(dx^{b}+a^{b}_t)+\tau^{-1}\left(e^{2\sigma}(dr^2+dz^2)-r^2dt^2\right)~~,~~\tau=-\det(\lambda_{ab})~~,
\eeq
and
\beq
A=\sqrt{3}\psi_\phi d\phi+\sqrt{3}\psi_\psi d\psi+A_t dt~~.
\eeq
The quantities $a^{t}_b$ and $\sigma$ are determined in terms of the fields $\Phi^{A}$ according to the relations presented in \cite{Tomizawa:2009ua,Armas:2009dd}. The electric potentials $\psi_a$, the magnetic potential $\mu$ and the twist potentials $\omega_a$ are determined via Einstein and Maxwell equations that follow from \eqref{eq:action} and obey the following relations,
\begin{eqnarray}
d\psi_a&=&-\frac{1}{\sqrt{3}}i_{V_{(a)}}F~~,\label{eq:electric}\\
d\mu&=&\frac{1}{\sqrt{3}}*\left(V_{(1)}\wedge V_{(2)}\wedge F\right)-\epsilon^{ab}\psi_a d\psi_b~~,\label{eq:magnetic}\\
d\omega_a&=&*\left(V_{(1)}\wedge V_{(2)}\wedge dV_{(a)}\right)+\psi_a\left(3d\mu+\epsilon^{bc}\psi_bd\psi_c\right) \label{eq:twist}~~,
\end{eqnarray}
where $\epsilon^{\phi\psi}=-\epsilon^{\psi\phi}=1$.

If we consider two different field configurations corresponding to two different solutions $\Theta_{0}$ and $\Theta_{1}$ one can define the deviation matrix $\Psi=\Theta_{1}\Theta_{0}^{-1}-1$, which measures the difference between the two solutions and vanishes if the two solutions are identical. Furthermore the matrix $\Theta$ defines a conserved current $J^{a}=\Theta^{-1}\partial^{a}\Theta$ for each solution, which can be used to define the difference between the conserved currents of two solutions $\bar{J}^{a}=\Theta^{-1}_1\partial^{a}\Theta_1-\Theta^{-1}_0\partial^{a}\Theta_0$. Due to the properties of the matrix $\Theta$ we can decompose it as $\Theta=\hat{g}\hat{g}^{T}$ where $\hat{g}$ is a $G_{2(2)}$ matrix. With this in hand we can define the matrix $M^{a}=\hat{g}_0\bar{J}^{a}T\hat{g}_{1}$ which measures the difference between the conserved currents of the two solutions and vanishes if the two conserved currents are identical. It is then possible to derive the Mazur identity for this sigma-model \cite{Tomizawa:2009ua}
\beq \label{eq:Mazur}
\int_{\partial\Sigma}r\partial_\mu\tr \Psi dS^{\mu}=\int_\Sigma r h_{\mu\nu}\tr\left(M^{T\nu}M^{\nu}\right)drdz~~,
\eeq
where $h_{\mu\nu}dx^{\mu}dx^{\nu}=dr^2+dz^2$. This identity expresses the fact that for two solutions of \eqref{eq:action} with isometry group $\mathbb{R}\times U(1) \times U(1)$ it is only necessary to show that the deviation matrix $\Psi$ vanishes on the boundary of the base space $\Sigma$ defined at $r=0$ and at infinity, since that implies, according to Eq.~\eqref{eq:Mazur}, that the solution is identical on the entire base space $\Sigma$. The proof then relies on the classification of the boundary $r=0$ according to its domain structure \cite{Harmark:2004rm, Harmark:2005vn, Harmark:2009dh,Armas:2011ed} and in the measurement of local charges and fluxes.

%%%%%%%%%%%%%%%%%%%%%%%%%%%%%%%%%%%%%%%%%%%%%%%%%%%%%%%%%%
\subsection{Domain structure of black hole space-times} \label{sec:domain}
Black hole space-times are characterised by their domain structure \cite{Harmark:2004rm, Harmark:2005vn, Harmark:2009dh,Armas:2011ed}. If the base space $\Sigma$ is two-dimensional then the domain structure for the theory \eqref{eq:action} reduces to the rod structure \cite{Harmark:2009dh, Armas:2009dd}, in which the $z$-axis representing $r=0$ is split into a set of rods (intervals) $I_i=(\kappa_i,\kappa_{i+1})$ characterised by the particular linear combination of the Killing vectors $V_{(0)},V_{(a)}$ that vanishes on that interval. Each interval has a specific length $l_{i}=\kappa_{i+1}-\kappa_i$ and each interval is associated with a rod vector $v$. For two adjacent space-like rods with rod vectors $v=m_1 V_{(1)}+n_1 V_{(2)}$ and $v'=m_2 V_{(1)}+n_2 V_{(2)}$ with $m_1,n_1,n_2,m_2\in \mathbb{Z}$ we must have that $m_1n_2-m_2n_1=\pm1$.

We consider a generic rod structure for black hole space-times which are asymptotically flat (see Sec.~\ref{sec:kaluza} for the case of asymptotically Kaluza-Klein) and posses three commuting Killing vector fields. Such space-times must have at least two semi-infinite rods describing the fixed planes of rotation associated with the $\phi$ and $\psi$ directions and, in addition, an asymptotic region. These are characterised by\footnote{We took the liberty of adding the boundary at infinity $\partial\Sigma_\infty$ to the rod structure. Strictly speaking, $\partial\Sigma_\infty$ is not part of the rod structure.}
\begin{itemize}
\item \textbf{(i)} semi-infinite $\phi$-invariant plane: $\partial\Sigma_\phi^{-}=\{(r,z)|r=0~,~-\infty<z<\kappa_1\}$ and rod vector $v=(0,1,0)$.
\item \textbf{(ii)} semi-infinite $\psi$-invariant plane: $\partial\Sigma_\psi^{+}=\{(r,z)|r=0~,~\kappa_l<z<\infty\}$ and rod vector $v=(0,0,1)$.
\item \textbf{(iii)} infinity: $\partial\Sigma_\infty=\{(r,z)|\sqrt{r^2+z^2}\to\infty~,~\frac{z}{\sqrt{r^2+z^2}}=constant\}$.
\end{itemize} 
These two semi-infinite rods are depicted in the figure below. 
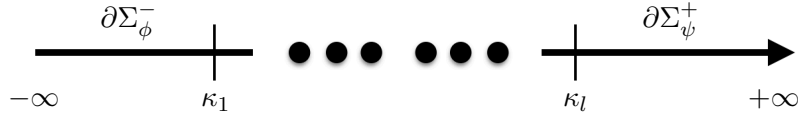
\begin{figure}[!ht]
\centerline{\includegraphics[scale=0.5]{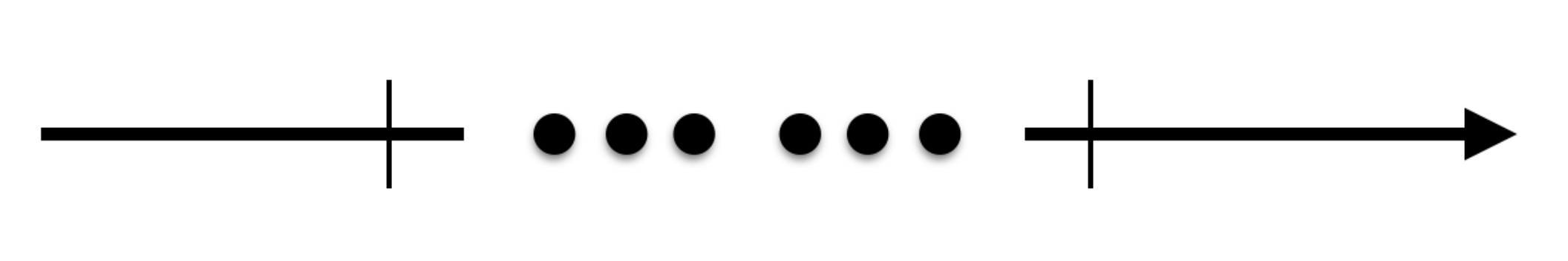} }
\begin{picture}(0,0)(0,0)
\put(90,20){ $-\infty   $}
\put(370,20){ $+\infty   $}
\put(163,20){ $\kappa_1   $}
\put(300,20){ $\kappa_l $}
\put(125,50){ $\partial\Sigma^{-}_{\phi}  $}
\put(330,50){ $\partial\Sigma^{+}_{\psi}  $}
\end{picture}		
\vskip -0.5cm
\caption{Generic rod structure diagram of asymptotically flat black holes in five dimensions. The \emph{dots} represent any possible arrangement of other spatial compact rods or horizon rods.} 
	\label{fig:1}
\end{figure}
In the middle of these two rods we can place in an arbitrary arrangement, an arbitrary number of rods of the following type:
\begin{itemize}
\item \textbf{(iv)} compact $\phi$-invariant plane: $\partial\Sigma_\phi=\{(r,z)|r=0~,~\kappa_i<z<\kappa_{i+1}\}$ and rod vector $v=(0,1,0)$.
\item \textbf{(v)} compact $\psi$-invariant plane: $\partial\Sigma_\psi=\{(r,z)|r=0~,~\kappa_i<z<\kappa_{i+1}\}$ and rod vector $v=(0,0,1)$.
\item \textbf{(vi)} compact space-like rod: $\partial\Sigma_\mathcal{B}=\{(r,z)|r=0~,~\kappa_i<z<\kappa_{i+1}\}$ and rod vector $v=(0,n,p)$.
\item \textbf{(vii)} black hole, black lens or black ring: $\partial\Sigma_{\mathcal{H}}=\{(r,z)|r=0~,~\kappa_i<z<\kappa_{i+1}\}$ with rod vector $v=(1,\Omega_\phi,\Omega_\psi)$ and, in the case of the black ring, where the $S^{1}$ is either parametrised by $\phi$ or by $\psi$. Furthermore, $\Omega_\phi$ and $\Omega_\psi$ are the angular velocities of the horizon.
\end{itemize}

The topology of the black hole depends on the other two adjacent rods. If the black hole rod \textbf{(vii)} is placed in between a rod of the type $\textbf{(iv)}$ and $\textbf{(v)}$ then it has spherical topology, whereas, if it is placed between two rods of type $\textbf{(iv)}$ or type $\textbf{(v)}$ it has $S^{1}\times S^{2}$ topology. Moreover, if it is placed between a rod of type $\textbf{(iv)}$ and another of type $\textbf{(vi)}$ it is a Lens space. Similarly, the topology of a rod of the type $\textbf{(vi)}$ depends on the adjacent rods. If it touches a rod of the type \textbf{(vii)} on one side and a rod of the type \textbf{(iv)} on the other it has the topology of a disk, while, if it is placed between two rods of the type \textbf{(iv)} or \textbf{(v)} it constitutes a non-trivial 2-cycle (bubble) with $S^{2}$ topology. We note that, despite the fact that the rods of the type \textbf{(vi)} include the rods \textbf{(iv)} and \textbf{(v)} as special cases, we have separately introduced them to connect with previous results in the literature and to highlight the differences between them. Furthermore we label the spatial compact rods of type \textbf{(iv)}-\textbf{(vii)} by $k$ and the horizon rods \textbf{(vii)} by $i$.

We will first focus on rod structures with one single connected horizon and then on the case of multiple disconnected horizons. Our strategy is to split the left hand side of Eq.~\eqref{eq:Mazur} as
\beq\label{eq:Mazurleft}
\int_{\partial\Sigma}r\partial_\mu\tr \Psi dS^{\mu}=\sum_I \int_I r\partial_z\tr \Psi dz+\int_{\partial\Sigma_\infty}r\partial_\mu\tr \Psi dS^{\mu}~~,
\eeq
and provide the fields $\Phi^{A}=\{\lambda_{ab},\omega_a,\psi_a,\mu\}$ with appropriate boundary conditions on each rod and at infinity. We will use \eqref{eq:Mazurleft} in order to elaborate a proof of uniqueness theorems for black holes with one single connected horizon in Sec.~\ref{sec:single}, with multiple disconnected horizons in Sec.~\ref{sec:multiple} and for asymptotically Kaluza-Klein black holes in Sec.~\ref{sec:kaluza}.

%%%%%%%%%%%%%%%%%%%%%%%%%%%%%%%%%%%%%%%%%%%%%%%%%%%%%%%%%%
\subsection{Boundary value problem for single connected horizon} \label{sec:single}
In this section we will show uniqueness of the most generic non-extremal black hole solution with one single connected horizon, in particular, we wish to give a proof of the following theorem,

\paragraph{Theorem 1:} \emph{In five-dimensional minimal supergravity, an asymptotically flat non-extremal black hole solution with a single connect horizon that is regular on and outside the event horizon is uniquely characterised by its rod structure, asymptotic charges (mass, angular momenta and electric charge) and magnetic fluxes if the black hole space-time admits, besides the stationary Killing vector, two mutually commuting axial Killing vector fields.} \\

The boundary value problem for each of the rods presented above has been analysed in detail in \cite{Tomizawa:2009ua, Tomizawa:2009tb,Armas:2009dd}, except for rods of the type \textbf{(vi)} which are responsible for the existence of non-trivial 2-cycles with generic rod vector $v=(0,n,p)$. We will begin by summarising the boundary value problem on the other rods.

\paragraph{Rods \textbf{(i)} and \textbf{(ii)}:} 
On these rods we have, according to Eqs.~\eqref{eq:magnetic}-\eqref{eq:twist} together with the fact that $\psi_a$ vanishes at infinity, that the magnetic potential and the twist potential are constant along the rod. Furthermore, according to Eq.~\eqref{eq:electric}, $\psi_\phi$ is constant on the $\phi$-invariant plane and $\psi_\psi$ is constant on the $\psi$-invariant plane. This results on conditions for $\lambda_{ab}$ and $\psi_{a}$ that do not have parameters that need to be specified while for the magnetic and twist potentials we have
\beq
\mu^{\pm}=c_0^{\pm}+\mathcal{O}(r^2)~,~\omega_a^{\pm}=c_a^{\pm}+\mathcal{O}(r^2)~.
\eeq
The six constants $c_0^{\pm},c_a^{\pm}$ can be fixed by looking at the integral of $\omega_a$ and $\mu$ over the entire $z$-axis and relating it to integrations at infinity via Stokes theorem. One concludes that \cite{Tomizawa:2009ua, Tomizawa:2009tb,Armas:2009dd}
\beq \label{eq:total}
c_0^{\pm}=\pm2\frac{Q^{T}}{\sqrt{3}\pi}~,~c_a^{\pm}=\pm\frac{2}{\pi}J^{T}_{a}~~,
\eeq
where $Q^{T}$ is the total electric charge of the solution and $J^{T}_{a}$ the total angular momentum of the solution associated with the $a=(\phi,\psi)$-fixed planes of rotation. The asymptotic charges $Q^{T}$ and $J^{T}_{a}$ will be defined in the next section. This completely specifies the solution in these rods and one can indeed check that their contribution to \eqref{eq:Mazurleft} vanishes. Even though no more quantities are necessary to be specified in these intervals, it is convenient to define the magnetic flux on these two rods over the surface $C^{\pm}$ as in \cite{Armas:2009dd}. This is given by 
\beq
\Phi^{\pm}=\int_{C^{\pm}} F~~.
\eeq
Integrating this explicitly we find $\Phi^{-}=2\pi\sqrt{3}\psi_{\psi}(\kappa_1)$ and $\Phi^{+}=-2\pi\sqrt{3}\psi_{\phi}(\kappa_N)$. This gives necessary boundary conditions for adjacent rods, namely, we have that
\beq \label{eq:psipsi}
(\psi_\phi,\psi_\psi)(\kappa_1)=(0,\frac{\Phi^{-}}{2\pi\sqrt{3}})~~\text{on $\partial\Sigma^{-}$}~~,~~(\psi_\phi,\psi_\psi)(\kappa_{N})=(-\frac{\Phi^{+}}{2\pi\sqrt{3}},0)~~\text{on $\partial\Sigma^{+}$}~~.
\eeq
In order to show uniqueness of spherical black holes and black rings (or of a spherical black hole with an arbitrary number of concentric rings) without the existence of 2-cycles, the fluxes \eqref{eq:psipsi} do not need to be specified \cite{Armas:2009dd}. On the other hand, if the horizon has Lens space topology it is necessary to specify one of them \cite{Armas:2009dd}. However, we will see that in the presence of 2-cycles they must in general be specified. 

\paragraph{Rod \textbf{(iii)}:} At infinity the vanishing of the second term on the right hand side of Eq.~\eqref{eq:Mazurleft} is guaranteed by the asymptotic falloffs of the fields. These falloffs are given in, e.g., Sec.~4.3 of \cite{Tomizawa:2009ua}. At $\partial\Sigma_\infty$ two solutions are identical if they have the same total mass $M$, angular momenta $J_a^{T}$ and electric charge $Q^{T}$ \cite{Tomizawa:2009ua}.

\paragraph{Rods \textbf{(iv)} and \textbf{(v)}:} On these rods the magnetic and twist potentials are not necessarily constant. The analysis is the same for both types of rod but for the sake of brevity we focus on the rod \textbf{(iv)}. In this case, from \eqref{eq:electric} we have that $d\psi_\phi=0$ and hence $\psi_\phi$ is constant over the rod while $\psi_\psi=f(z)$ for some function $f(z)$. Using \eqref{eq:magnetic}-\eqref{eq:twist} one finds that \cite{Tomizawa:2009tb}
\beq \label{eq:f}
\omega_\psi=d_\psi-\psi_\phi f^2(z)~,~\omega_\phi=d_\phi-2\psi_\phi^2f(z)~,~\mu=d_\mu-\psi_\phi f(z)~~,
\eeq 
where the constants $d_\psi,d_\phi,d_\mu,\psi_\phi$ must be specified in order to conclude that the contribution of this rod to \eqref{eq:Mazurleft} vanishes, while the function $f(z)$ needs not. However, in order to determine $d_\psi,d_\phi,d_\mu,\psi_\phi$ one needs to know $f(z)$ in general at one of the endpoints of the rod. If we imagine this rod to be adjacent to \textbf{(i)}, then according to \eqref{eq:psipsi} and continuity of the potentials one has that $\psi_\phi=0$ on $\partial\Sigma_\phi$ in which case $f(z)$ does not play a role while $d_\psi=c_\psi^{-},d_\phi=c_\phi^-,d_\mu=c_0^{-}$. If instead the rod was adjacent to \textbf{(ii)} then in order to determine the four constants one uses the fact that from \eqref{eq:psipsi} we have that $\psi_\psi(\kappa_N)=f(\kappa_N)=0$ and that $\psi_\phi=-(2\pi\sqrt{3})^{-1}\Phi^{+}$. However, in the case that the rod \textbf{(iv)} would be adjacent to another rod \textbf{(iv)} meeting at $z=\kappa_{i+1}$ we would require knowledge of $f(\kappa_{i+1})$. To obtain this we define the magnetic flux on the rod over the surface $C_\phi$ such that 
\beq \label{eq:phiphi}
\Phi_{\phi}=\int_{C_\phi}F=2\pi\sqrt{3}\left(\psi_\psi(\kappa_{i+1})-\psi_{\psi}(\kappa_i)\right)~~,
\eeq
and similarly for $\Phi_\psi$ on $\partial\Sigma_\psi$. Therefore, given the measurement of $\Phi_\phi$ we can obtain the value of $\psi_\psi(\kappa_{i+1})$ and consequently of all the remaining potentials using \eqref{eq:f} at $z=\kappa_{i+1}$ given $\psi_\psi(\kappa_{i})$.

Summarizing, given the knowledge of the potentials $\omega_\psi,\omega_\phi,\psi_\phi,\mu,\psi_\psi$ at one endpoint $\kappa_i$ of the rod, using \eqref{eq:f} and \eqref{eq:phiphi}, the values of these potentials are obtained at the other endpoint $\kappa_{i+1}$ of the rod and hence provide boundary conditions for any other rod of the type \textbf{(iv)}-\textbf{(vii)} that may be adjacent to it. Since we are now considering only the case of one single connected horizon, any rod of the type \textbf{(iv)}-\textbf{(v)} can only be adjacent to (at most) one horizon rod. Therefore, since we have previously determined all the values of the potentials in the rods \textbf{(i)}-\textbf{(ii)} via \eqref{eq:total} and \eqref{eq:psipsi}, and now also the change in the potentials across the rods \textbf{(iv)}-\textbf{(v)} via \eqref{eq:f} and \eqref{eq:phiphi}, we only need to show that this can also be done for rods of the type \textbf{(vi)}. This is now analysed below.

\paragraph{Rods \textbf{(vi)}:} In this case we will give the precise details of how to derive the boundary conditions on the various fields since the inclusion of this type of rods is new. It turns out that finding the correct boundary conditions in this case is very similar to the case analysed in \cite{Tomizawa:2009tb, Armas:2009dd} where the rod vector was taken to be $v=(0,1,p)$. We consider a rod vector of the type $v=(0,n,p)$ which by definition corresponds to the Killing vector field $V=n\partial_\phi+p\partial_\psi$ with fixed points on $\partial\Sigma_\mathcal{B}$. Therefore, we have that $g(v,v)=0$ and that $\tau$, defined in \eqref{eq:metric}, vanishes on $\partial\Sigma_\mathcal{B}$. This implies the following relations
\beq
n^2\lambda_{\phi\phi}+2np\lambda_{\phi\psi}+p^2\lambda_{\psi\psi}=0~~,~~\lambda_{\phi\phi}\lambda_{\psi\psi}-\lambda_{\phi\psi}^2=0~~.
\eeq
Assuming $n$ to be non-zero and defining $k=p/n$, from here we deduce that
\beq 
\lambda_{\phi\phi}\sim k^2g(z)+\mathcal{O}(r^2)~~,~~\lambda_{\phi\psi}\sim- kg(z)+\mathcal{O}(r^2)~~,~~\lambda_{\psi\psi}\sim g(z)+\mathcal{O}(r^2)~~,
\eeq
for some function $g(z)$ that needs not to be specified. Furthermore, from \eqref{eq:electric} we have that $i_v F=nd\psi_\phi+pd\psi_\psi=0$ and hence integrating we get $\psi_\phi=d_0-k\psi_\psi$. Therefore we have that
\beq \label{eq:f1}
\psi_\phi\sim d_0-kf(z)+\mathcal{O}(r^2)~~,~~\psi_\psi\sim f(z)+\mathcal{O}(r^2)~~,
\eeq
for some function $f(z)$. Using now Eqs.~\eqref{eq:magnetic}-\eqref{eq:twist} we find
\beq \label{eq:f2}
\mu=	-d_0f(z)+d_1+\mathcal{O}(r^2)~~,~~\omega_\phi=-2d_0^2f(z)+kd_0f^2(z)+d_2+\mathcal{O}(r^2)~~,~~\omega_\psi=-d_0f^2(z)+d_3+\mathcal{O}(r^2)~~. 
\eeq
The vanishing of this rod's contribution to \eqref{eq:Mazurleft} requires that all four constants $d_0,d_1,d_2,d_3$ are specified. As in the previous case of the rods \textbf{(iv)} and \textbf{(v)} we have the same number of constants which can be determined by knowing what the potentials $\omega_a,\mu,\psi_\phi$ and $f(z)$ are at one endpoint of the rod $z=\kappa_i$. When only one horizon rod is present this is always possible since we know all the values of the potentials at the endpoints of the rods \textbf{(i)} and \textbf{(ii)}. Therefore we can keep on updating their value according to \eqref{eq:f}, \eqref{eq:f1},\eqref{eq:f2} by following the two chains of rods, which begin at the leftmost and rightmost rods, until we hit the horizon rod. Since this type of rods \textbf{(vi)} can be adjacent to rods of the same type or to \textbf{(iv)} and \textbf{(v)}, it is necessary to know how the potentials $\psi_a$ vary across the rod and hence determine all the potentials at $z=\kappa_{i+1}$ so that they provide boundary conditions for the adjacent rods. This is done by defining the magnetic flux through the surface $C_{\mathcal{B}}$ such that
\beq\label{eq:phib}
\begin{split}
\Phi_{\mathcal{B}}=\int_{C_{\mathcal{B}}}F=&\thinspace2\pi\sqrt{3}(1-k)\left(\psi_\psi(\kappa_{i+1})-\psi_\psi(\kappa_{i})\right)\\
=&\thinspace2\pi\sqrt{3}\frac{(k-1)}{k}\left(\psi_\phi(\kappa_{i+1})-\psi_\phi(\kappa_{i})\right)~~.
\end{split}
\eeq
Therefore, given $\Phi_{\mathcal{B}}$ we can determine $\psi_\phi(\kappa_{i+1})$ and $\psi_\psi(\kappa_{i+1})$. If this rod is placed in between two other compact space-like rods then the surface $C_{\mathcal{B}}$ has the topology of an $S^{2}$ and this rod represents a non-trivial 2-cycle. As mentioned previously, this type of rods includes the cases of the rods \textbf{(iv)} and \textbf{\textbf{(v)}}, in particular, if we set $n=1$ and $p=0$ in \eqref{eq:phib} we obtain \eqref{eq:phiphi}. We define the set of magnetic fluxes on the spatial compact rods by $\Phi^{k}=\{\Phi_a,\Phi_{\mathcal{B}}\}$ and the set of all magnetic fluxes as $\Phi_M=\{\Phi^{\pm},\Phi^{k}\}$.

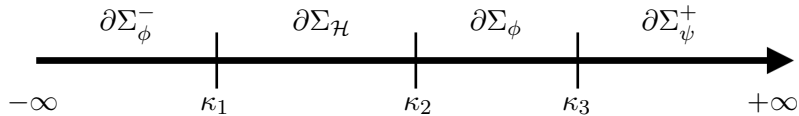
\begin{figure}[!ht]
\centerline{\includegraphics[scale=0.5]{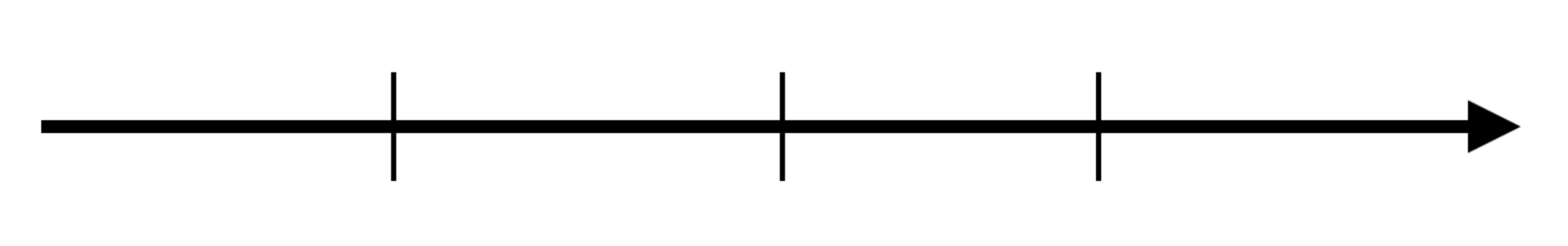} }
\begin{picture}(0,0)(0,0)
\put(90,20){ $-\infty   $}
\put(370,20){ $+\infty   $}
\put(163,20){ $\kappa_1   $}
\put(300,20){ $\kappa_3 $}
\put(240,20){ $\kappa_2 $}
\put(125,50){ $\partial\Sigma^{-}_{\phi}  $}
\put(330,50){ $\partial\Sigma^{+}_{\psi}  $}
\put(265,50){ $\partial\Sigma_{\phi}  $}
\put(198,50){ $\partial\Sigma_\mathcal{H}  $}
\end{picture}		
\vskip -0.5cm
\caption{Rod structure diagram for an asymptotically flat black ring in five dimensions.} 
	\label{fig:2}
\end{figure}

\paragraph{Rods \textbf{(vii)}:} In the case of an horizon rod we simply need that $\lambda_{ab}\sim\mathcal{O}(1)~,~\psi_{a}\sim\mathcal{O}(1)~,~\mu\sim\mathcal{O}(1)~,~\omega_{a}\sim\mathcal{O}(1)$ for their contribution to \eqref{eq:Mazurleft} to vanish \cite{Tomizawa:2009ua}. However, this means that as soon as we cross an horizon in the middle of a rod structure diagram, we loose all the information regarding the change in the potentials $\psi_a,\mu,\omega_a$. In the case where we only consider one single horizon rod, this does not constitute a problem since, as mentioned previously, we follow the two chains of rods starting from the left and right of the horizon rod. In the case where multiple horizons are present it requires measuring extra charges as it will be explained in the next section. We note, however, that in the case of the rod \textbf{(vii)} representing a black ring horizon, it is possible to measure its dipole charge via the formula
\beq \label{eq:dipole}
q_{\psi}=\frac{1}{2\pi}\int_{S^{2}}F=\sqrt{3}\left(\psi_{\phi}(\kappa_{i+1})-\psi_\phi(\kappa_i)\right)~~,
\eeq
where the $S^{2}$ encloses the ring once and where we have assumed the $S^{1}$ to be parametrised by $\psi$. Analogously, we can define the same type of charge for a ring with $S^{1}$ parametrised by $\phi$. In same cases, specifying this charge is enough to show uniqueness without having to measure the magnetic fluxes $\Phi^{\pm}$. However, measuring the fluxes $\Phi^{\pm}$ is always sufficient for all the solutions and the dipole charge is not needed. For example, consider the simplest case of a black ring analysed in \cite{Tomizawa:2009tb} and with the rod structure depicted in Fig.~\ref{fig:2}. 
According to \eqref{eq:psipsi} we have that $\psi_\phi(\kappa_1)=0$ on $\partial\Sigma^{-}_{\phi}$ and hence, according to \eqref{eq:dipole} we have that $\psi_\phi(\kappa_2)=(\sqrt{3})^{-1}q_\psi$. Since we have that $\psi_\phi$ is constant on $\partial\Sigma_{\phi}$ then we deduce that $\psi_\phi(\kappa_3)=(\sqrt{3})^{-1}q_\psi$. Now, again due to \eqref{eq:psipsi} we find that there is a non-trivial flux $\Phi^{+}=-2\pi q_\psi$. Therefore, we see that the presence of the dipole charge induces a non-trivial flux $\Phi^{+}$ on $\partial\Sigma_\psi^{+}$. This means that to show uniqueness of the black ring solution we can instead require the measurement of $\Phi^{+}$ and not of $q_\psi$, suggesting that, since measuring $\Phi^{\pm}$ is always necessary in the presence of arbitrary large sequences of bubbles, the fluxes $\Phi^{\pm}$ are physically more relevant charges than $q_\psi,q_\phi$.  To summarize, generically the magnetic fluxes $\Phi^{\pm}$ obey the following relations
\beq \label{eqs:dipole}
\begin{split}
\int_{\partial\Sigma_{\mathcal{H}}}d\psi_\phi&=-\frac{1}{2\pi\sqrt{3}}\left(\Phi^{-}+\sum_{I_{\phi}}\Phi_{\phi}+\sum_{I_{\mathcal{B}}}\Phi_{\mathcal{B}}\right)~~,\\
\int_{\partial\Sigma_{\mathcal{H}}}d\psi_\psi&=-\frac{1}{2\pi\sqrt{3}}\left(\Phi^{-}+\sum_{I_{\psi}}\Phi_{\psi}+\sum_{I_{\mathcal{B}}}\Phi_{\mathcal{B}}\right)~~,\\
\end{split}
\eeq
where $I_{\phi}$, $I_{\psi}$ and $I_{\mathcal{B}_p}$ label all the intervals associated with the rods \textbf{(iv)}, \textbf{(v)} and \textbf{(vi)} respectively. In the case that the horizon is a black ring than the left hand side of both equalities above is just the dipole charge of the black ring up to a factor of $\sqrt{3}$ as defined in \eqref{eq:dipole}.

This ends the proof of the uniqueness theorem for a single connected horizon component leading us to conclude that a black hole space-time with a single horizon and three commuting Killing vector fields is uniquely characterised by its rod structure and the set of charges and fluxes $\{M,J_{a}^T,Q^T,\Phi_{M}\}$. Below, we focus on the case where multiple disconnected horizons may be present.

%%%%%%%%%%%%%%%%%%%%%%%%%%%%%%%%%%%%%%%%%%%%%%%%%%%%%%%%%%
\subsection{Boundary value problem for multiple disconnected horizons} \label{sec:multiple}
As mentioned previously, in the case of multiple disconnected horizons it is necessary to specify further charges measured locally near each horizon. To understand this better consider the following rod arrangement in which a rod vector of the type \textbf{(iv)} is placed in between two rods of the type \textbf{(vii)} as depicted in the figure below.
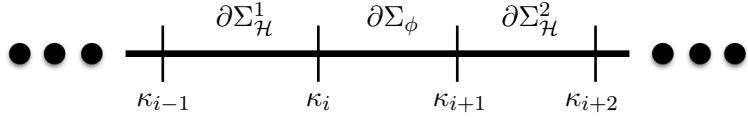
\begin{figure}[!ht]
\centerline{\includegraphics[scale=0.5]{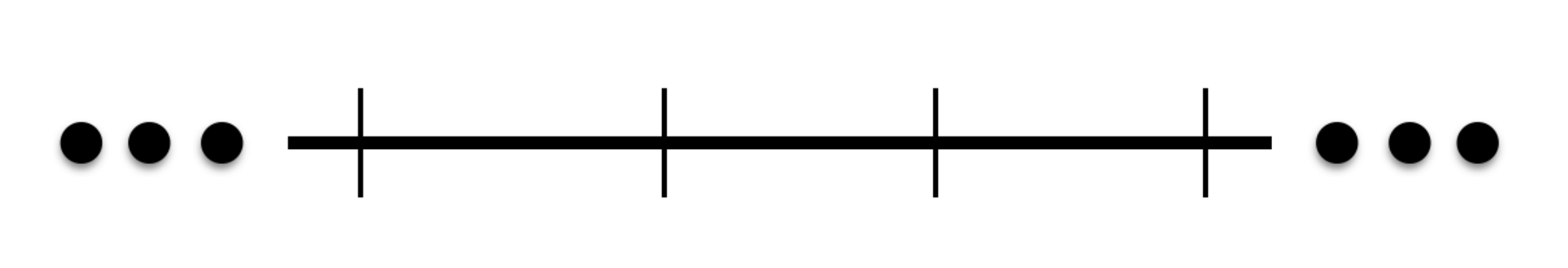} }
\begin{picture}(0,0)(0,0)
\put(153,20){ $\kappa_{i-1}   $}
\put(265,20){ $\kappa_{i+1} $}
\put(315,20){ $\kappa_{i+2} $}
\put(217,20){ $\kappa_{i} $}
\put(183,50){ $\partial\Sigma_{\mathcal{H}}^{1}   $}
\put(240,50){ $\partial\Sigma_{\phi}  $}
\put(291,50){ $\partial\Sigma_{\mathcal{H}}^{2}$}
\end{picture}		
\vskip -0.5cm
\caption{An example of a rod diagram with two consecutive horizons separated by a compact space-like rod.} 
	\label{fig:3}
\end{figure}
 In order to provide boundary conditions for the rod \textbf{(iv)} and determine the constants in \eqref{eq:f} we need to know how the potentials $\omega_a,\mu$ varied across the horizon rod $i$. As explained in \cite{Armas:2009dd}, this requires the measurement of the angular momenta and electric charge near each horizon, namely,\footnote{Note that the expression for $J_{a}^{i}$ in \eqref{eq:charges} is equivalent to $J_{a}^{i}=(\pi/4)\int_{\Sigma_{\mathcal{H}}}\left(*\left(V_{(1)}\wedge V_{(2)}\wedge dV_{(a)}\right)+\psi_a\left(3d\mu+\epsilon^{bc}\psi_bd\psi_c\right) \right)~~$ as presented in \cite{Armas:2009dd}.}
\beq \label{eq:charges}
\begin{split}
J_{a}^{i}=&\frac{1}{16\pi}\int_{\mathcal{H}}*dV_{(a)}+\frac{\sqrt{3}}{16\pi}\int_{\mathcal{H}}\psi_a\left(*F+\frac{2}{\sqrt{3}}A\wedge F\right)~~, \\
Q^{i}=&\frac{1}{16\pi}\int_{\mathcal{H}}\left(*F+\frac{1}{\sqrt{3}}A\wedge F\right)~~.
\end{split}
\eeq
In the first line of \eqref{eq:charges} we have defined the angular momenta measured near each horizon and we note that the first term is the usual Komar integral on the horizon while the second term is the electromagnetic contribution to the angular momenta. If the integration is taken over $\partial\Sigma_{\infty}$ instead, then these integrals define the total charges $J^{T}_{a}$ and $Q^{T}$ introduced in \eqref{eq:total}. We note that in the case of $J^{T}_{a}$, the contribution due to the second term in \eqref{eq:charges} vanishes at infinity \cite{Tomizawa:2009ua}. These charges are just an integrated version of the right hand side of  Eqs.~\eqref{eq:magnetic}-\eqref{eq:twist}. Therefore we have the change in potentials $\mu(\kappa_{i})-\mu(\kappa_{i-1})=4(\sqrt{3}\pi)^{-1}Q^{i}$ and $\omega_a(\kappa_{i})-\omega_a(\kappa_{i-1})=(4/\pi)J_a^{i}$.

However, the charges \eqref{eq:charges} do not provide any information regarding the change in the potentials $\psi_a$ across the horizon rods and these are required in order to determine the constants involved in \eqref{eq:f}. In the case that one of the two horizon rods is a black ring then the dipole charge \eqref{eq:dipole} gives us the value of $\psi_\phi$ on the rod since $d\psi_\phi=0$ there. However, in order to determine the remaining constants $d_\psi,d_\phi ,d_\mu$ it is necessary to know $f(z)$ at least at one endpoint of the rod $\partial\Sigma_{\phi}$. The question that remains is if it is possible to acquire knowledge of $f(z)$ at one of the endpoints $z=\kappa_i$ or $z=\kappa_{i+1}$ just by local measurements on that rod. In fact, this question was answered in \cite{Armas:2009dd} and requires the measurement of the flux $\Phi_{\phi}$ on $\partial\Sigma_{\phi}$ introduced in \eqref{eq:phiphi} as well as the Chern-Simons flux measured on $\partial\Sigma_{\phi}$ defined as
\beq \label{eq:xi}
\Xi_{\phi}=\int_{C_\phi}A\wedge(i_{V_{(2)}} F)=-3\pi\left(\psi_\psi^2(\kappa_{i+1})-\psi_\psi^2(\kappa_{i})\right)~~,
\eeq
and similarly $\Xi_{\psi}$ in the $\psi$-invariant plane. The flux \eqref{eq:xi} together with the flux \eqref{eq:phiphi} allows us to find a unique solution to $f(\kappa_i)$ given by \cite{Armas:2009dd}
\beq \label{eq:fi}
f(\kappa_i)=-\frac{1}{\sqrt{3}}\left(\frac{\Xi_\phi}{\Phi_\phi}+\frac{\Phi_\phi}{4\pi}\right)~~.
\eeq
With a unique solution to $f(\kappa_i)$ we can find the constants $d_\psi,d_\phi ,d_\mu$ provided that we have determined $\psi_\phi$ via the dipole charge of one of the horizons rods. However, we could image the two horizon rods in Fig.~\ref{fig:3} to represent two spherical black holes, even though such configurations will unlikely be regular. In such case, we cannot define a dipole charge over the horizon and hence cannot determine in this way the value of $\psi_\phi$ in the adjacent rod between the two horizons. Nevertheless, we can, from \eqref{eq:fi} and \eqref{eq:phiphi} or \eqref{eq:xi}, deduce the value of $f(\kappa_{i+1})$, which when used in any of the equations \eqref{eq:f} allows us to determine $\psi_\phi$ in general. Therefore, in this case one does not have to specify the dipole charge of the black ring horizons.

If one now replaces the middle rod in Fig.~\ref{fig:3} by one of the type \textbf{(vi)} then one can ask the same question, namely, if local measurements on the rod can determine the value of $f(\kappa_i)$. Indeed, by using \eqref{eq:phib} and defining the analog of \eqref{eq:xi},
\beq \label{eq:xib}
\Xi_{\mathcal{B}}=\int_{C_\mathcal{B}}A\wedge(i_{V_{(2)}} F)=-3\pi(1-k)\left(\psi_\psi^2(\kappa_{i+1})-\psi_\psi^2(\kappa_i)\right)-\frac{\sqrt{3}}{(1-k)}d_0\Phi_{\mathcal{B}}~~,
\eeq
one obtains\footnote{Note that in \eqref{eq:xib} we have defined the flux $\Xi_{\mathcal{B}}$ with respect to the Killing vector $V_{(2)}$. We could have defined it also with respect to $V_{(1)}$, which would result in a similar expression for $f(\kappa_i)$.}
\beq \label{eq:fib}
f(\kappa_i)=-\frac{(1-k)}{\sqrt{3}}\left(\frac{\Xi_{\mathcal{B}}}{\Phi_{\mathcal{B}}}+\frac{\Phi_{\mathcal{B}}}{4\pi(1-k)^2}+\frac{\sqrt{3}}{(1-k)}d_0\right)~~.
\eeq
Using this result into Eqs.~\eqref{eq:f2} one can determine all the constants $d_0,d_1,d_2,d_3$ as previously for the rod $\partial\Sigma_{\phi}$ depicted in Fig.~\ref{fig:3}. We further note that \eqref{eq:xib} includes \eqref{eq:xi} as a special case when $n=1$ and $p=0$. Hence we define the set of fluxes on the spatial compact rods as $\Xi_M=\{\Xi_{a},\Xi_{\mathcal{B}}\}$.

\paragraph{Problemata:} The unique solutions of $f(\kappa_i)$ presented in \eqref{eq:fi} and in \eqref{eq:fib} are not valid when $\Phi_\phi=0$ and $\Phi_{\mathcal{B}}=0$. In such cases we can only conclude that $f(\kappa_i)=f(\kappa_{i+1})$ and that the fluxes $\Xi_M$ and $\Phi^{k}$ do not  determine the solution in these rods. This problem has also been addressed in \cite{Armas:2009dd} and its solution requires specifying two other set of charges in the case that at least one of the two horizons depicted in Fig.~\ref{fig:3} is a black ring. In that case, we can define the Maxwell charge $Q_M^{i}$ and a Chern-Simons dipole charge $\mathcal{Q}_\psi^{i}$ measured near each black ring horizon via\footnote{In theories with Chern-Simons terms there are different notions of charges \cite{Marolf:2000cb}. The electric charge defined in \eqref{eq:charges} is also known as the Page charge.}
\beq \label{eq:maxwell}
Q_M^{i}=\frac{1}{16\pi}\int_{\mathcal{H}}*F~~,~~\mathcal{Q}_\psi^{i}=\frac{1}{2\pi}\int_{S^{2}}A\wedge (i_{V_{(2)}}F)~~,
\eeq
where in order to perform the $\mathcal{Q}_\psi^{i}$ integration one needs to specify, as in the case of the dipole charge \eqref{eq:dipole}, a tangent vector along the ring. Using this into the definition of $Q^{i}$ in \eqref{eq:charges} together with \eqref{eq:magnetic} we find the recursive relation \cite{Armas:2009dd}
\beq
\psi_{\phi}(\kappa_{i+1})f(\kappa_{i+1})=\psi_{\phi}(\kappa_{i+2})\psi_\psi(\kappa_{i+2})-\frac{4}{\sqrt{3}\pi}\left(Q^{i}-Q^{i}_M\right)-\frac{2}{3}\mathcal{Q}^{i}_\psi~~.
\eeq
Therefore, we can determine the function $f(\kappa_{i+1})$ across the horizon of a black ring if we measure its dipole charges and Maxwell charge \eqref{eq:dipole},\eqref{eq:maxwell}. In fact, in the cases where at least one horizon represents a black ring then specifying $q_a^{i},\mathcal{Q}_a^{i},Q_M^{i}$ yields the desired value of $f(\kappa_{i+1})$ even when the fluxes $\Phi^{k}$ are not given, but of course, at the expense of measuring another local charge on the black ring horizon.\footnote{Note that we are labelling the dipole charge $q_a$ introduced in \eqref{eq:dipole} measured near each black ring horizon $i$ by $q^{i}_a$.} Note that in general, the fluxes $\Phi_{\pm}$ also obey the relations \eqref{eqs:dipole}, where now the left hand side should be modified to a sum over all horizon rods. In the case that the horizons are all black rings then we can relate the integral over each horizon to its dipole charge \eqref{eq:dipole} or Chern-Simons dipole charge \eqref{eq:maxwell}.

Summarizing, if the magnetic flux $\Phi_a$ or $\Phi_\mathcal{B}$ on the single rod in between the two horizons depicted in Fig.~\ref{fig:3} vanish and one of the horizons is not a black ring, we cannot show uniqueness of such configurations. More generally, consider a combination of the rods of the following form,
\begin{figure}[!ht]
\centerline{\includegraphics[scale=0.5]{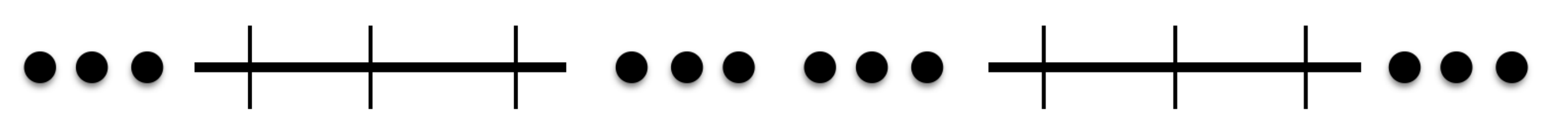} }
\begin{picture}(0,0)(0,0)
\put(117,42){ $\Sigma_{\mathcal{H}}^{1}   $}
\put(149,42){ $\partial\Sigma_{1}  $}
\put(318,42){ $\partial\Sigma_{2}  $}
\put(353,42){ $\Sigma_{\mathcal{H}}^{2}  $}
\end{picture}		
\vskip -0.5cm
\caption{An example of a rod diagram with two consecutive horizons separated by an arbitrary series of compact space-like rods.} 
	\label{fig:4}
\end{figure}
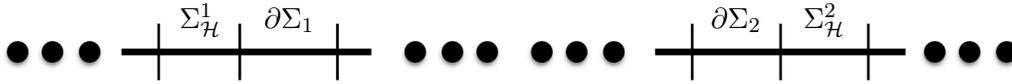
where $\partial\Sigma_1$ or $\partial\Sigma_2$ can be any type of spatial compact rod, while the \emph{dots} represent any type and number of spatial compact rods. Furthermore assume that none of the horizons is a black ring, then, if the magnetic flux $\Phi_a$ or $\Phi_\mathcal{B}$ vanish on $\partial\Sigma_1$ and simultaneously on $\partial\Sigma_2$, we cannot show uniqueness. \\

With this in mind we define the two sets of constraints, 
\begin{itemize}
\item \textbf{(1)} The fluxes $\Phi_a$ or $\Phi_\mathcal{B}$ on any two adjacent rods $\partial\Sigma_1$ or $\partial\Sigma_2$ to two consecutive horizons $\Sigma_{\mathcal{H}}^{1}$ and $\Sigma_{\mathcal{H}}^{2}$ as in Fig.~\ref{fig:4} do not vanish simultaneously.
\item \textbf{(2)} In an arbitrary arrangement of rods, only one of the horizon rods $\Sigma_{\mathcal{H}}$ has Lens-space or spherical topology while the remaining have $S^{1}\times S^{2}$ topology.
\end{itemize}
With these constraints defined we have proven the following theorem:

\paragraph{Theorem 2:} \emph{Consider, in five-dimensional minimal supergravity, an asymptotically flat non-extremal black hole solution with multiple disconnected horizons that is regular on and outside each of the event horizons which admits, besides the stationary Killing vector, two mutually commuting axial Killing vector fields. Then if the constraints \textbf{(1)} are realised, the black hole space-time is uniquely characterised by its rod structure, charges and fluxes $\{M,J^{T}_{a},Q^{T},J_{a}^{i},Q^{i},\Phi_{M},\Xi_{M} \}$, and if the constraints \textbf{(2)} are realised, the black hole space-time is uniquely characterised by its rod structure, charges and fluxes $\{M,J^{T}_{a},Q^{T},J_{a}^{i},Q^{i},\Phi_{M},q_a^{i}$ $,\mathcal{Q}_a^{i},Q_{M}^{i} \}$ .} \\

We will now show how the previous two theorems can be generalised to the context of asymptotically Kaluza-Klein black holes.

%%%%%%%%%%%%%%%%%%%%%%%%%%%%%%%%%%%%%%%%%%%%%%%%%%%%%%%%%%
\subsection{Asymptotically Kaluza-Klein black holes} \label{sec:kaluza}
In this section we generalise the theorems of Secs.~\ref{sec:single} and \ref{sec:multiple} to the case of black holes which are asymptotically Kaluza-Klein. Uniqueness theorems for such black holes in the case of a single connected horizon and in the absence of non-trivial 2 cycles were studied in \cite{Tomizawa:2010xj}. The difference between Kaluza-Klein black holes and the asymptotically flat black holes studied in the previous sections resides on the different leftmost and rightmost rods. As depicted in the figure below, these two rods have the following characteristics,
\begin{itemize}
\item \textbf{(viii)} leftmost semi-infinite plane: $\partial\Sigma^{-}=\{(r,z)|r=0~,~-\infty<z<\kappa_1\}$ and rod vector $v=(0,1,-N)$.
\item \textbf{(ix)} rightmost semi-infinite plane: $\partial\Sigma^{+}=\{(r,z)|r=0~,~\kappa_l<z<\infty\}$ and rod vector $v=(0,1,N)$.
\end{itemize}
Here $N$ stands for the NUT charge of the asymptotically Kaluza-Klein black hole. The boundary of $\Sigma$ for Kaluza-Klein black holes also contains an asymptotic region of the form \textbf{(iii)}. 
\begin{figure}[!ht]
\centerline{\includegraphics[scale=0.5]{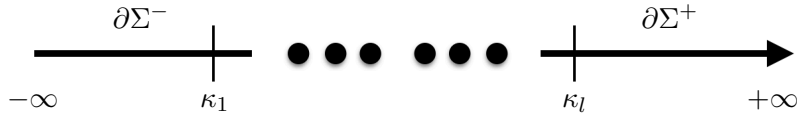} }
\begin{picture}(0,0)(0,0)
\put(90,20){ $-\infty   $}
\put(370,20){ $+\infty   $}
\put(163,20){ $\kappa_1   $}
\put(300,20){ $\kappa_l $}
\put(130,50){ $\partial\Sigma^{-}  $}
\put(330,50){ $\partial\Sigma^{+}$}
\end{picture}		
\vskip -0.5cm
\caption{Generic rod structure diagram of asymptotically flat Kaluza-Klein black holes in five dimensions. The \emph{dots} represent any possible arrangement of other spatial compact rods or horizon rods.} 
	\label{fig:5}
\end{figure}
For the contribution of this region to the integral \eqref{eq:Mazurleft} to vanish, it is necessary that the mass $M$, the angular moment $J_a^{T}$, the electric charge $Q^T$, the NUT charge $N$ and the magnetic flux defined as
\beq
\Phi_{\infty}=\int_{S^2_{\infty}}F~~,
\eeq
where $S^2_{\infty}$ denotes the base manifold of $S^{2}$ at infinity, are the same for both solutions \cite{Tomizawa:2010xj}.
In between these two rods depicted in Fig.~\ref{fig:5} we can place the same type of rods as those analysed in the previous sections. The analysis is then the same as long as we guarantee that we have all the information about the potentials $\psi_a,\mu,\omega_a$ at $z=\kappa_1$ and $z=\kappa_l$, hence providing the necessary boundary conditions for any rod that may be adjacent to it. The boundary value problem for the rods \textbf{(viii)} and \textbf{(ix)} is the same as for the rod \textbf{(vi)} where now we have $n=1,p=\pm N$ and hence $k=\pm N$. Focusing on the case of the rod \textbf{(viii)} (the case of the rod \textbf{(ix)} is essentially the same apart from minus signs), we see from \eqref{eq:f1}-\eqref{eq:f2} that we must have
\beq \label{eq:psik}
\psi_\phi\sim d_0-Nf(z)+\mathcal{O}(r^2)~~,~~\psi_\psi\sim f(z)+\mathcal{O}(r^2)~~,
\eeq
for some function $f(z)$ and
\beq
\mu=	-d_0f(z)+d_1+\mathcal{O}(r^2)~~,~~\omega_\phi=-2d_0^2f(z)+Nd_0f^2(z)+d_2+\mathcal{O}(r^2)~~,~~\omega_\psi=-d_0f^2(z)+d_3+\mathcal{O}(r^2)~~. 
\eeq
By looking at the asymptotic behaviour of the fields at infinity one finds that \cite{Tomizawa:2010xj}
\beq
d_0=-\frac{\Phi_\infty}{\sqrt{3}}~~,~~d_1=\frac{2Q^T}{\pi L}~~,~~d_2=-\frac{3}{4\pi L}J^{T}_{\phi}+\frac{1}{\pi L}\left(\Phi_{\infty}Q^{T}-NJ_{\psi}\right)~~,~~d_3=\frac{3}{\pi L}J_{\psi}^{T}~~,
\eeq
where $2\pi L$ is the periodicity of the fifth Kaluza-Klein dimension and we have used the fact that $f(\pm \infty)=0$. In order to determine all the potentials at $z=\kappa_1$ one requires the knowledge of $f(\kappa_1)$. This is obtained by defining an analogous flux to \eqref{eq:phib}
\beq
\begin{split}
\Phi_{N}^{-}=\int_{C_{N}^{-}}F=&\thinspace2\pi\sqrt{3}(1+N)\left(\psi_\psi(\kappa_{1})-\psi_\psi(-\infty)\right)\\
=&\thinspace2\pi\sqrt{3}(1+N)f(\kappa_{1})~~,
\end{split}
\eeq
where we have used \eqref{eq:psik} and that $f(-\infty)=0$. Defining the flux $\Phi_{N}^{+}$ in the rightmost rod completes the proof of the uniqueness theorem for asymptotically Kaluza-Klein black holes and generalises the two theorems stated in the previous two sections where the fluxes $\Phi_{N}^{\pm}$ now take the role of the fluxes $\Phi^{\pm}$ and in addition one needs to further specify the NUT charge $N$ and the magnetic flux at infinity $\Phi_\infty$.

%%%%%%%%%%%%%%%%%%%%%%%%%%%%%%%%%%%%%%%%%%%%%%%%%%%%%%%%%%
\section{Discussion} \label{sec:conclusions}
In this note we have generalised previous uniqueness theorems for asymptotically flat and asymptotically Kaluza-Klein black holes with multiple disconnected horizons and with three mutually independent Killing vector fields with isometry group $\mathbb{R}\times U(1)\times U(1)$ \cite{Tomizawa:2009ua, Tomizawa:2009tb,Armas:2009dd, Tomizawa:2010xj} to include the cases where arbitrary sequences of non-trivial 2-cycles may be present in the domain of outer communication. In the case in which a single horizon is present in the rod structure diagram, the theorem that we have proved includes all possible rod combinations allowed by the rod structure constraints (see Sec.~\ref{sec:domain}). We have shown that such black hole space-times are uniquely characterised by their rod structure, asymptotic charges $\{M,J^{T}_a,Q^{T}\}$ and a set of magnetic fluxes $\Phi_{M}=\{\Phi^{\pm},\Phi_a,\Phi_{\mathcal{B}}\}$ defined on the semi-infinite and compact spatial rods of the rod diagram, which depending on the rod arrangement have either $S^{2}$ (bubble) or disk topology. In particular, we noted that given the magnetic fluxes $\Phi_{M}$ and in the case that the horizon corresponds to a black ring, there was no need to specify its dipole charge. In the case of Kaluza-Klein black holes, the set of asymptotic charges also includes the NUT charge $N$ and the magnetic flux measured at infinity $\Phi_\infty$. We note that in general, such as in the presence of multiple disconnected horizons or non-trivial 2-cycles, the charges at infinity are not sufficient to uniquely describe the solution. In addition, magnetic fluxes, dipole charges, angular momenta and electric charges measured near each horizon must also be specified. 

In the case of multiple disconnected horizons we have provided a generalisation of \cite{Armas:2009dd}, however we have not been able to show uniqueness of the most general class of solutions, hypothetically allowed by the rod structure formalism. In between any two horizons of arbitrary topology, one can define magnetic fluxes $\Phi^{k}$ on each of the spatial compact rods adjacent to each of the horizons. If these fluxes do not vanish simultaneously on each of the adjacent rods then we have shown uniqueness of the most general solution, in which case one needs to specify the local angular momenta $J_{a}^{i}$ and local electric charge $Q^{i}$ of each horizon besides the Chern-Simons fluxes $\Xi_M$ on each spatial compact rod. However, if the two fluxes $\Phi_M$ vanish simultaneously, then we have only been able to shown uniqueness of a class solutions with an arbitrary arrangement of rods as long as only one horizon with either Lens-space or spherical topology is present in the rod diagram while the remaining have $S^{1}\times S^{2}$ topology. In this latter case, in order to classify the black hole space-time uniquely, one has to specify the Maxwell charge $Q_M^{i}$, the dipole charge $q_a^{i}$ and the Chern-Simons dipole charge $\mathcal{Q}_a^{i}$ of each of the black ring horizons, instead of specifying the Chern-Simons fluxes $\Xi_M$ on each spatial compact rod. 

The domain structure of black holes provides a set of geometrical and topological invariants for any black hole space-time \cite{Harmark:2004rm, Harmark:2005vn, Harmark:2009dh,Armas:2011ed}. However, even though it sets some constraints on the allowed solutions, it does not give information regarding whether or not a specific arrangement of domains (or rods) may actually be realised as a regular solution in a given theory. When considering the most general solution with multiple disconnected horizons in Sec.~\ref{sec:multiple}, we have allowed for any possible combination of the rods presented in Sec.~\ref{sec:domain}, including, for example, an infinite series of spherically rotating black holes or Lens-spaces. However, such solutions are known to be hard to realise, free of conical singularities, at least in pure Einstein gravity \cite{Evslin:2008gx, Chen:2008fa, Herdeiro:2009zz}. This suggests that perhaps the fact that it seems difficult to classify the most generic solution may be an indication that such solutions could not be regular. Further constraints on the solutions, from regularity conditions, may be obtained by developing inverse scattering methods, along the lines of \cite{Figueras:2009mc, Compere:2009zh}, that can deal with non-trivial 2-cycles. The reason why we have not been able to perform this most generic classification is because across the rod of an event horizon which does not have $S^{1}\times S^{2}$ topology we cannot define a magnetic or electric flux over a space-time surface that relates the values of the electric potentials $\psi_a$ at the two rod endpoints. Such type of fluxes for an horizon with $S^{1}\times S^{2}$ topology can be defined and are essentially its dipole charges. The question of whether or not another set of charges can be defined for horizons with other topologies may be answered by generalising the work of \cite{Kunduri:2013vka} in order to understand what charges may enter the first law of thermodynamics of black hole objects with multiple disconnected horizons in minimal supergravity. We note that alternatively, one may consider no longer insisting that black holes must be specified by certain 'physical' charges, but instead generically by certain integrals over the rod diagrams. This of course would solve the problem of quantifying the change in the electric potentials $\psi_a$ across the horizon of a spherical black hole.

The domain structure provides us with a set of invariants such as the domain areas or rod lengths. For a given black hole space-time, these lengths are in general related to the physical charges of the black hole or have to be adjusted in a particular way in order to yield a space-time free of conical singularities. When showing such uniqueness theorems based on the rod structure, it is still an open question whether or not the total amount of information is over-specifying the solution. With this in mind, it would be interesting to understand generically what information is contained in each rod interval. For example, the length of the horizon rod is known to contain information about the entropy of the horizon \cite{Hollands:2008fm} but perhaps more information can be extracted from the space-like rods. 

One can consider extending the theorems presented here to extremal black holes. In such cases it is known that the near-horizon topologies have to be classified and shown to be unique \cite{Figueras:2009ci}. In the context of the theory \eqref{eq:action} this has still not been fully done though many have been classified \cite{Kunduri:2009ud, Kunduri:2011zr, Kunduri:2013gce}. While it may turn out that in the extremal case, uniqueness theorems may be harder to prove, at least when these black holes are non-extremal and have two rotational symmetries, we have given a very satisfactory answer, in particular, it includes a possible non-extremal version of the black hole constructed in \cite{Kunduri:2014iga}.

\section*{Acknowledgements}
JA is thankful to Matthias Blau, Adolfo Guarino and Troels Harmark for useful discussions. JA is specially grateful to Hari K. Kunduri and James Lucietti for useful e-mail correspondence. JA would also like to thank Jorge V. Rocha and two anonymous referees for excellent comments on this manuscript. JA would like to thank the organisers of the conference \textbf{Recent Developments in String Theory} in Ascona (2014) for a lovely time. This work has been supported by the Swiss National Science Foundation and the �Innovations- und Kooperationsprojekt C-13� of the Schweizerische Universit\"{a}tskonferenz SUK/CUS.

\hypersetup{urlcolor=blue}

%%%%%%%%%%%%%%%%%%%%%%%%%%%%%%%%%%%%%%%%%%%%%%%%%%%%%%%%%%%%%
\addcontentsline{toc}{section}{References}
\footnotesize
\providecommand{\href}[2]{#2}\begingroup\raggedright\endgroup


\begin{thebibliography}{10}

\bibitem{Gibbons:2013tqa}
G.~Gibbons and N.~Warner, ``{Global structure of five-dimensional fuzzballs},''
  \href{http://dx.doi.org/10.1088/0264-9381/31/2/025016}{{\em
  Class.Quant.Grav.} {\bf 31} (2014)  025016},
\href{http://arxiv.org/abs/1305.0957}{{\tt arXiv:1305.0957 [hep-th]}}.
%%CITATION = ARXIV:1305.0957;%%.

\bibitem{Kunduri:2013vka}
H.~K. Kunduri and J.~Lucietti, ``{The first law of soliton and black hole
  mechanics in five dimensions},''
  \href{http://dx.doi.org/10.1088/0264-9381/31/3/032001}{{\em
  Class.Quant.Grav.} {\bf 31} (2014)  032001},
\href{http://arxiv.org/abs/1310.4810}{{\tt arXiv:1310.4810 [hep-th]}}.
%%CITATION = ARXIV:1310.4810;%%.

\bibitem{Kunduri:2014iga}
H.~K. Kunduri and J.~Lucietti, ``{Black hole non-uniqueness via spacetime
  topology in five dimensions},''
\href{http://arxiv.org/abs/1407.8002}{{\tt arXiv:1407.8002 [hep-th]}}.
%%CITATION = ARXIV:1407.8002;%%.

\bibitem{Bena:2005va}
I.~Bena and N.~P. Warner, ``{Bubbling supertubes and foaming black holes},''
  \href{http://dx.doi.org/10.1103/PhysRevD.74.066001}{{\em Phys.Rev.} {\bf D74}
  (2006)  066001},
\href{http://arxiv.org/abs/hep-th/0505166}{{\tt arXiv:hep-th/0505166
  [hep-th]}}.
%%CITATION = HEP-TH/0505166;%%.

\bibitem{Mathur:2005zp}
S.~D. Mathur, ``{The Fuzzball proposal for black holes: An Elementary
  review},'' \href{http://dx.doi.org/10.1002/prop.200410203}{{\em
  Fortsch.Phys.} {\bf 53} (2005)  793--827},
\href{http://arxiv.org/abs/hep-th/0502050}{{\tt arXiv:hep-th/0502050
  [hep-th]}}.
%%CITATION = HEP-TH/0502050;%%.

\bibitem{Mathur:2008nj}
S.~D. Mathur, ``{Fuzzballs and the information paradox: A Summary and
  conjectures},''
\href{http://arxiv.org/abs/0810.4525}{{\tt arXiv:0810.4525 [hep-th]}}.
%%CITATION = ARXIV:0810.4525;%%.

\bibitem{Harmark:2004rm}
T.~Harmark, ``{Stationary and axisymmetric solutions of higher-dimensional
  General Relativity},''
  \href{http://dx.doi.org/10.1103/PhysRevD.70.124002}{{\em Phys. Rev.} {\bf
  D70} (2004)  124002},
\href{http://arxiv.org/abs/hep-th/0408141}{{\tt arXiv:hep-th/0408141}}.
%%CITATION = HEP-TH/0408141;%%.

\bibitem{Harmark:2005vn}
T.~Harmark and P.~Olesen, ``{On the structure of stationary and axisymmetric
  metrics},'' \href{http://dx.doi.org/10.1103/PhysRevD.72.124017}{{\em
  Phys.Rev.} {\bf D72} (2005)  124017},
\href{http://arxiv.org/abs/hep-th/0508208}{{\tt arXiv:hep-th/0508208
  [hep-th]}}.
%%CITATION = HEP-TH/0508208;%%.

\bibitem{Harmark:2009dh}
T.~Harmark, ``{Domain Structure of Black Hole Space-Times},''
  \href{http://dx.doi.org/10.1103/PhysRevD.80.024019}{{\em Phys. Rev.} {\bf
  D80} (2009)  024019},
\href{http://arxiv.org/abs/0904.4246}{{\tt arXiv:0904.4246 [hep-th]}}.
%%CITATION = 0904.4246;%%.

\bibitem{Armas:2011ed}
J.~Armas, P.~Caputa, and T.~Harmark, ``{Domain Structure of Black Hole
  Space-Times with a Cosmological Constant},''
  \href{http://dx.doi.org/10.1103/PhysRevD.85.084019}{{\em Phys.Rev.} {\bf D85}
  (2012)  084019},
\href{http://arxiv.org/abs/1111.1163}{{\tt arXiv:1111.1163 [hep-th]}}.
%%CITATION = ARXIV:1111.1163;%%.

\bibitem{Tomizawa:2009ua}
S.~Tomizawa, Y.~Yasui, and A.~Ishibashi, ``{A uniqueness theorem for charged
  rotating black holes in five-dimensional minimal supergravity},''
  \href{http://dx.doi.org/10.1103/PhysRevD.79.124023}{{\em Phys. Rev.} {\bf
  D79} (2009)  124023},
\href{http://arxiv.org/abs/0901.4724}{{\tt arXiv:0901.4724 [hep-th]}}.
%%CITATION = 0901.4724;%%.

\bibitem{Tomizawa:2009tb}
S.~Tomizawa, Y.~Yasui, and A.~Ishibashi, ``{Uniqueness theorem for charged
  dipole rings in five-dimensional minimal supergravity},''
  \href{http://dx.doi.org/10.1103/PhysRevD.81.084037}{{\em Phys.Rev.} {\bf D81}
  (2010)  084037},
\href{http://arxiv.org/abs/0911.4309}{{\tt arXiv:0911.4309 [hep-th]}}.
%%CITATION = ARXIV:0911.4309;%%.

\bibitem{Armas:2009dd}
J.~Armas and T.~Harmark, ``{Uniqueness Theorem for Black Hole Space-Times with
  Multiple Disconnected Horizons},''
  \href{http://dx.doi.org/10.1007/JHEP05(2010)093}{{\em JHEP} {\bf 05} (2010)
  093},
\href{http://arxiv.org/abs/0911.4654}{{\tt arXiv:0911.4654 [hep-th]}}.
%%CITATION = 0911.4654;%%.

\bibitem{Hollands:2012xy}
S.~Hollands and A.~Ishibashi, ``{Black hole uniqueness theorems in higher
  dimensional spacetimes},''
  \href{http://dx.doi.org/10.1088/0264-9381/29/16/163001}{{\em
  Class.Quant.Grav.} {\bf 29} (2012)  163001},
\href{http://arxiv.org/abs/1206.1164}{{\tt arXiv:1206.1164 [gr-qc]}}.
%%CITATION = ARXIV:1206.1164;%%.

\bibitem{Yazadjiev:2010uu}
S.~S. Yazadjiev, ``{A Uniqueness theorem for black holes with Kaluza-Klein
  asymptotic in 5D Einstein-Maxwell gravity},''
  \href{http://dx.doi.org/10.1103/PhysRevD.82.024015}{{\em Phys.Rev.} {\bf D82}
  (2010)  024015},
\href{http://arxiv.org/abs/1002.3954}{{\tt arXiv:1002.3954 [hep-th]}}.
%%CITATION = ARXIV:1002.3954;%%.

\bibitem{Tomizawa:2010xj}
S.~Tomizawa, ``{Uniqueness theorems for Kaluza-Klein black holes in
  five-dimensional minimal supergravity},''
  \href{http://dx.doi.org/10.1103/PhysRevD.82.104047}{{\em Phys.Rev.} {\bf D82}
  (2010)  104047},
\href{http://arxiv.org/abs/1007.1183}{{\tt arXiv:1007.1183 [hep-th]}}.
%%CITATION = ARXIV:1007.1183;%%.

\bibitem{Hollands:2007qf}
S.~Hollands and S.~Yazadjiev, ``{A Uniqueness theorem for 5-dimensional
  Einstein-Maxwell black holes},''
  \href{http://dx.doi.org/10.1088/0264-9381/25/9/095010}{{\em Class. Quant.
  Grav.} {\bf 25} (2008)  095010},
\href{http://arxiv.org/abs/0711.1722}{{\tt arXiv:0711.1722 [gr-qc]}}.
%%CITATION = 0711.1722;%%.

\bibitem{Marolf:2000cb}
D.~Marolf, ``{Chern-Simons terms and the three notions of charge},''
\href{http://arxiv.org/abs/hep-th/0006117}{{\tt arXiv:hep-th/0006117
  [hep-th]}}.
%%CITATION = HEP-TH/0006117;%%.

\bibitem{Evslin:2008gx}
J.~Evslin, ``{Geometric Engineering 5d Black Holes with Rod Diagrams},''
  \href{http://dx.doi.org/10.1088/1126-6708/2008/09/004}{{\em JHEP} {\bf 09}
  (2008)  004},
\href{http://arxiv.org/abs/0806.3389}{{\tt arXiv:0806.3389 [hep-th]}}.
%%CITATION = 0806.3389;%%.

\bibitem{Chen:2008fa}
Y.~Chen and E.~Teo, ``{A rotating black lens solution in five dimensions},''
  \href{http://dx.doi.org/10.1103/PhysRevD.78.064062}{{\em Phys. Rev.} {\bf
  D78} (2008)  064062},
\href{http://arxiv.org/abs/0808.0587}{{\tt arXiv:0808.0587 [gr-qc]}}.
%%CITATION = 0808.0587;%%.

\bibitem{Herdeiro:2009zz}
C.~A. Herdeiro, C.~Rebelo, M.~Zilhao, and M.~S. Costa, ``{A double Myers-Perry
  black hole: An inverse scattering construction},''
\href{http://dx.doi.org/10.1063/1.3141299}{{\em AIP Conf.Proc.} {\bf 1122}
  (2009)  296--299}.
%%CITATION = APCPC,1122,296;%%.

%\cite{Figueras:2009mc}
\bibitem{Figueras:2009mc} 
  P.~Figueras, E.~Jamsin, J.~V.~Rocha and A.~Virmani,
  %``Integrability of Five Dimensional Minimal Supergravity and Charged Rotating Black Holes,''
  Class.\ Quant.\ Grav.\  {\bf 27}, 135011 (2010)
  [arXiv:0912.3199 [hep-th]].
  %%CITATION = ARXIV:0912.3199;%%
  %28 citations counted in INSPIRE as of 14 Nov 2014
  
  %\cite{Compere:2009zh}
\bibitem{Compere:2009zh} 
  G.~Compere, S.~de Buyl, E.~Jamsin and A.~Virmani,
  %``G2 Dualities in D=5 Supergravity and Black Strings,''
  Class.\ Quant.\ Grav.\  {\bf 26}, 125016 (2009)
  [arXiv:0903.1645 [hep-th]].
  %%CITATION = ARXIV:0903.1645;%%
  %31 citations counted in INSPIRE as of 14 Nov 2014


\bibitem{Hollands:2008fm}
S.~Hollands and S.~Yazadjiev, ``{A uniqueness theorem for stationary
  Kaluza-Klein black holes},''
\href{http://arxiv.org/abs/0812.3036}{{\tt arXiv:0812.3036 [gr-qc]}}.
%%CITATION = 0812.3036;%%.

\bibitem{Figueras:2009ci}
P.~Figueras and J.~Lucietti, ``{On the uniqueness of extremal vacuum black
  holes},''
\href{http://arxiv.org/abs/0906.5565}{{\tt arXiv:0906.5565 [hep-th]}}.
%%CITATION = 0906.5565;%%.

\bibitem{Kunduri:2009ud}
H.~K. Kunduri and J.~Lucietti, ``{Static near-horizon geometries in five
  dimensions},''
\href{http://arxiv.org/abs/0907.0410}{{\tt arXiv:0907.0410 [hep-th]}}.
%%CITATION = 0907.0410;%%.

\bibitem{Kunduri:2011zr}
H.~K. Kunduri and J.~Lucietti, ``{Constructing near-horizon geometries in
  supergravities with hidden symmetry},''
  \href{http://dx.doi.org/10.1007/JHEP07(2011)107}{{\em JHEP} {\bf 1107} (2011)
   107},
\href{http://arxiv.org/abs/1104.2260}{{\tt arXiv:1104.2260 [hep-th]}}.
%%CITATION = ARXIV:1104.2260;%%.

\bibitem{Kunduri:2013gce}
H.~K. Kunduri and J.~Lucietti, ``{Classification of near-horizon geometries of
  extremal black holes},'' \href{http://dx.doi.org/10.12942/lrr-2013-8}{{\em
  Living Rev.Rel.} {\bf 16} (2013)  8},
\href{http://arxiv.org/abs/1306.2517}{{\tt arXiv:1306.2517 [hep-th]}}.
%%CITATION = ARXIV:1306.2517;%%.

\end{thebibliography}
\end{document}